\documentclass[conference]{IEEEtran}
\IEEEoverridecommandlockouts
\usepackage{cite}
\usepackage{amsmath,amsfonts}
\usepackage{graphicx}
\usepackage{subfigure}
\usepackage{tabularx}
\usepackage{multirow}

\begin{document}

\title{RAI-Net: Range-Adaptive LiDAR Point Cloud Frame Interpolation Network\\
\thanks{{*}Jianwen Chen is the corresponding author. © 2021 IEEE. Personal use of this material is permitted. Permission from IEEE must be obtained for all other uses, in any current or future media, including reprinting/republishing this material for advertising or promotional purposes, creating new collective works, for resale or redistribution to servers or lists, or reuse of any copyrighted component of this work in other works.}
}

\author{\IEEEauthorblockN{Lili Zhao, Zezhi Zhu, Xuhu Lin, Xuezhou Guo, Qian Yin, Wenyi Wang, and Jianwen Chen{*}}
\IEEEauthorblockA{\textit{School of Information and Communication Engineering} \\
\textit{University of Electronic Science and Technology of China}\\
Chengdu, China \\
chenjianwen@uestc.edu.cn}
}

\maketitle

\begin{abstract}
LiDAR point cloud frame interpolation, which synthesizes the intermediate frame between the captured frames, has emerged as an important issue for many applications. Especially for reducing the amounts of point cloud transmission, it is by predicting the intermediate frame based on the reference frames to upsample data to high frame rate ones. However, due to high-dimensional and sparse characteristics of point clouds, it is more difficult to predict the intermediate frame for LiDAR point clouds than videos. In this paper, we propose a novel LiDAR point cloud frame interpolation method, which exploits range images (RIs) as an intermediate representation with CNNs to conduct the frame interpolation process. Considering the inherited characteristics of RIs differ from that of color images, we introduce spatially adaptive convolutions to extract range features adaptively, while a high-efficient flow estimation method is presented to generate optical flows. The proposed model then warps the input frames and range features, based on the optical flows to synthesize the interpolated frame. Extensive experiments on the KITTI dataset have clearly demonstrated that our method consistently achieves superior frame interpolation results with better perceptual quality to that of using state-of-the-art video frame interpolation methods. The proposed method could be integrated into any LiDAR point cloud compression systems for inter prediction.
\end{abstract}


\section{Introduction}
\label{sec:intro}

Nowadays, \emph{light detection and ranging} (LiDAR) sensors are gaining more and more popularity in practical applications, such as self-driving vehicles, mobile robots, drones, tablet PC or mobile phones (e.g., iPhone 12 Pro). LiDAR can acquire the precise 3D digital representation of the surrounding environment, i.e., LiDAR point clouds, which consist of 3D coordinates indicating the locations of points. 
It is well-known that LiDAR point clouds have been widely utilized in many fields, e.g., augmented reality (AR), virtual reality (VR), simultaneous localization and mapping (SLAM)~\cite{SuMa2019}. 

LiDAR point cloud frame interpolation (i.e., temporal interpolation), which aims to synthesize the intermediate frame between the captured frames, begins to attract more and more attention, since it is greatly demanded in various tasks such as LiDAR point cloud comprssion~\cite{Tu_ACCESS}, frame rate up-conversion for data fusion~\cite{FrameRFusion}. LiDAR point cloud frame interpolation could address the above-mentioned issues by improving the frame rate. However, few works have eyes on it. Recently, PointINet~\cite{2020arXiv201210066L} is proposed for LiDAR point cloud frame interpolation, which is by estimating bi-directional 3D scene flow between the two point clouds and generating the interpolated frame based on the 3D scene flow. Note that the whole processing is conducted in 3D space with high complexity. This paper will focus on frame interpolation of LiDAR point clouds with low complexity.


\begin{figure}[t]
	\centering
	\includegraphics[width=83mm]{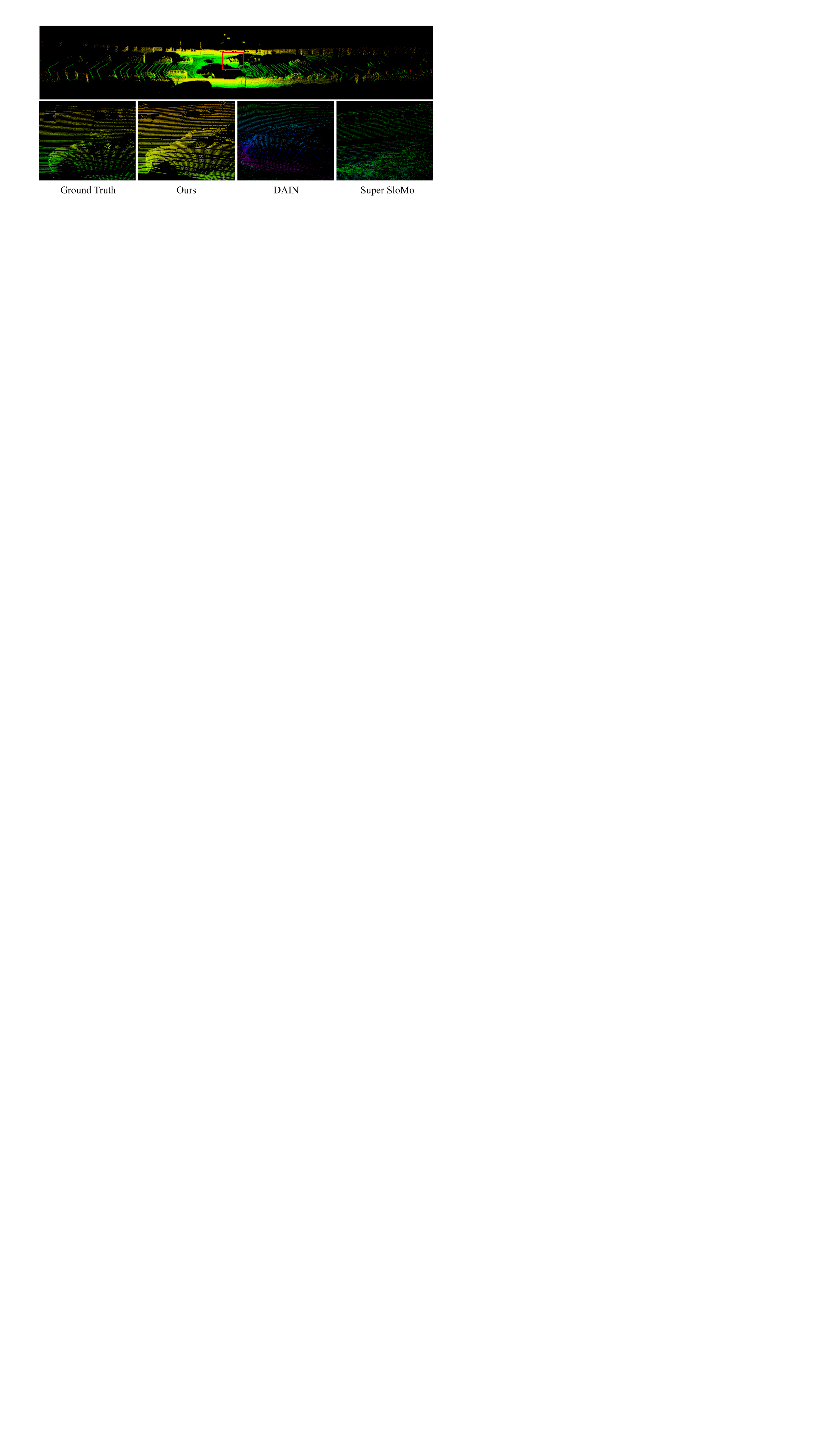}
	\caption{\textbf{An example for LiDAR point cloud frame interpolation.} We propose a range-adaptive point cloud interpolation approach. Our approach produces a high-quality result.}
	\label{fig:f1}
\end{figure}

Recent years have witnessed great success in applying deep convolutional neural networks (CNNs) on video frame interpolation (VFI). The existing approaches for VFI can be divided into three categories: phase-based methods (e.g.,~\cite{PhaseNet}), kernel-based methods (e.g.,~\cite{SepConv}), and flow-based methods (e.g.,~\cite{bao2019depth,jiang2018super,Softmax}). It can be observed that flow-based approaches perform well in quantitative benchmarks~\cite{2011}. Among them, the work DAIN~\cite{bao2019depth} via the advanced flow estimation method PWC-Net~\cite{sun2018pwc}, delivers the state-of-the-art performance. It is highly desirable to bring this technology to point cloud frame interpolation. However, it is not a direct extension. That is because point clouds have the characteristics: with the high dimensionality (i.e., 3D), huge amounts of data and the unstructured nature, imposing restrictions on adopting well-developed VFI methods (typically using 2D CNNs) directly.

\begin{figure}[t]
\centering
\subfigure[]{
\begin{minipage}[t]{3.4cm}
\centering
\includegraphics[width=3.4cm]{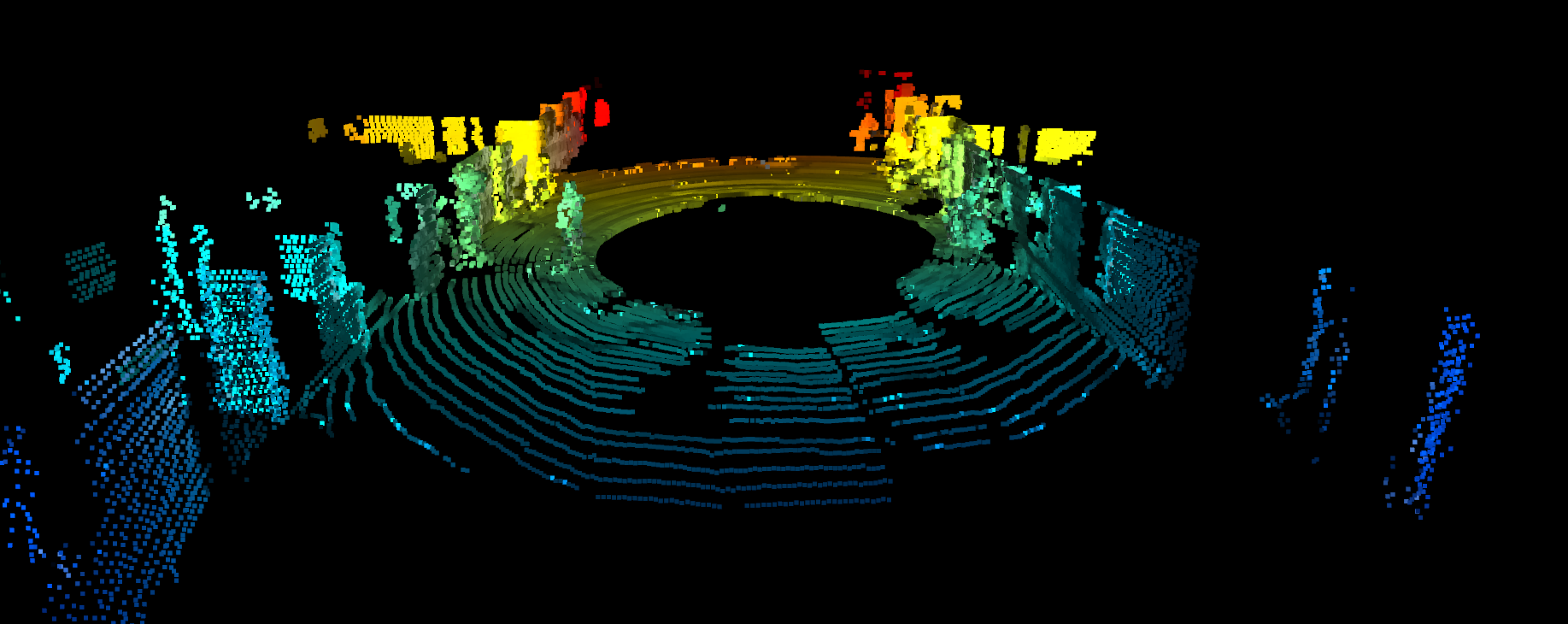}
\end{minipage}}
\subfigure[]{
\begin{minipage}[t]{4.4cm}
\centering
\includegraphics[width=4.4cm]{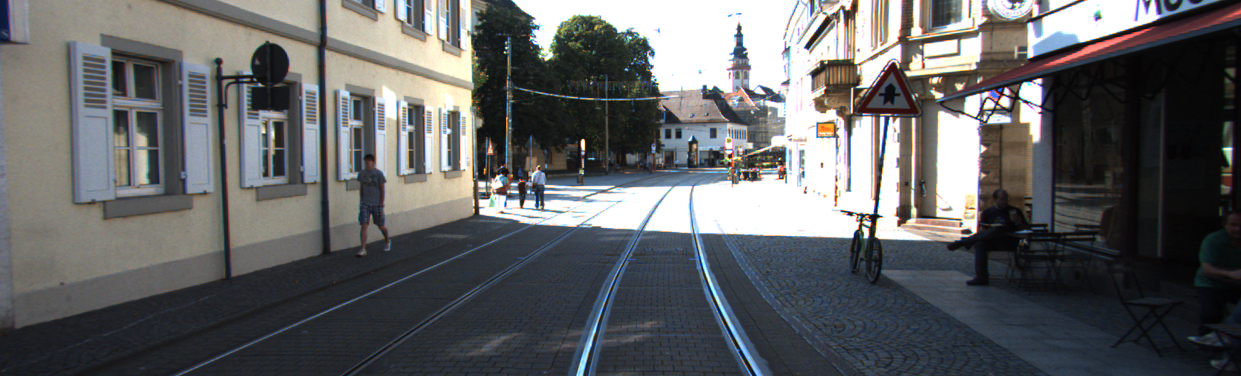}
\end{minipage}}
\subfigure[]{
\begin{minipage}[t]{8cm}
\centering
\includegraphics[width=8cm]{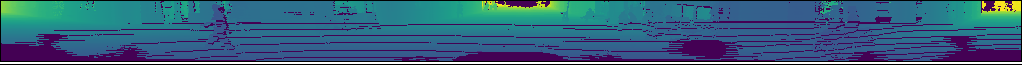}
\end{minipage}}
\caption{\textbf{Demonstrations of one frame from the KITTI dataset.} (a) The 3D LiDAR point cloud; (b) The corresponding RGB front-view image for reference; (c) The 2D range image projected from (a).}
\label{fig:example}
\end{figure}

It is worth noting that several works have explored low-complexity 3D segmentation via 2D representations of point clouds (e.g.,~\cite{RangeNet,xu2020squeezesegv3}). Instead of processing 3D points directly, point clouds are firstly projected into a set of 2D range images (RIs) [one can see Fig.~\ref{fig:example} (c)] and then segmented. Note that each pixel of RIs contains the range from each point to the center of the sensor's coordinate system. It is natural to consider this projection method for our work, where point clouds are transformed into RIs and then fed into VFI models to generate the interpolated frame. Nevertheless, naively applying these techniques usually leads to sub-optimal frame interpolation performance. It is likely due to the fact that the inherited characteristics of RIs differ from the normal color images. For example, the feature distribution of the color images is basically constant at their different spatial locations, while RIs are in the opposite case~\cite{xu2020squeezesegv3}. Meanwhile, the feature extractors adopted in the existing VFI approaches are typically standard convolutions, which have the spatial sharing nature, i.e., the parameters of convolutions are shared across the whole input. Therefore, the feature extractors may not be effective to handle the RIs, leading to relatively low frame interpolation performance.

\begin{figure}[t]
	\centering
	\includegraphics[width=85mm]{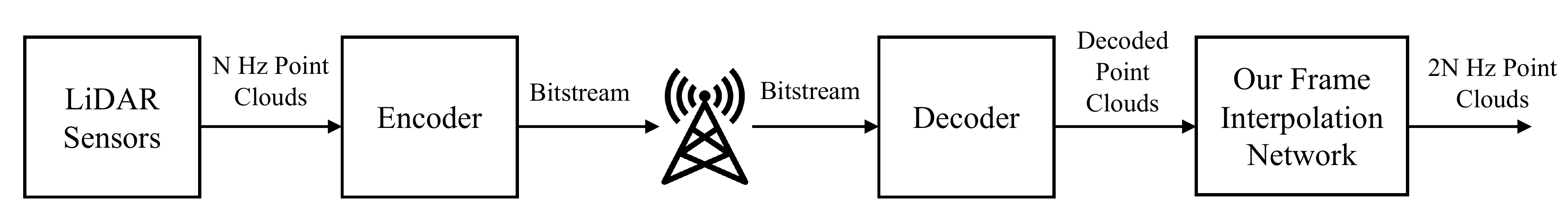}
	\caption{The illustration of the application into reducing the amount of the point cloud transmission.}
	\label{fig:application}
\end{figure}

In this paper, we address these challenges and propose a range-adaptive frame interpolation model, which can be applied to reduce the amounts of point cloud transmission (as shown in Fig.~\ref{fig:application}). The proposed approach follows the successful paths of flow-based VFI methods with novel adaptions. Specifically, first, 3D point clouds are projected to a set of 2D range images as input frames. Then, we estimate the bi-directional optical flows from two input frames via our flow estimator, which is modified from PWC-Lite, a lightweight version of PWC-Net~\cite{sun2018pwc}. Instead of the feature extractor adopted in the existing VFI methods for color images, we propose a range-adaptive feature extractor to obtain high-efficient representations of range images. Then, we utilize softmax splatting~\cite{Softmax} to warp the input frames and range features with the estimated flows. Finally, the in-between frame is generated. As shown in Fig.~\ref{fig:f1}, our method can deliver a competitive performance for point cloud frame interpolation.

The main contributions of this paper are the following: 
\begin{itemize}
	\item We propose a frame interpolation network for LiDAR point cloud, which is conducted by exploring range images as an intermediate representation combined with CNNs.
	\item To fully exploit spatial-temporal feature of point clouds as auxiliary information to synthesize the in-between frame, we introduce a range-adaptive feature extractor to produce spatial feature representations. On the other hand, we present a flow estimator for range images to generate high-efficient temporal feature.
	\item We demonstrate that the proposed method can deliver a favorable performance against state-of-the-art video frame interpolation models (i.e., DAIN~\cite{bao2019depth}, Super SloMo~\cite{jiang2018super}).
\end{itemize}

\section{Proposed Method}
\subsection{Algorithm Overview}
\begin{figure*}[t]
	\centering
	\includegraphics[width=180mm]{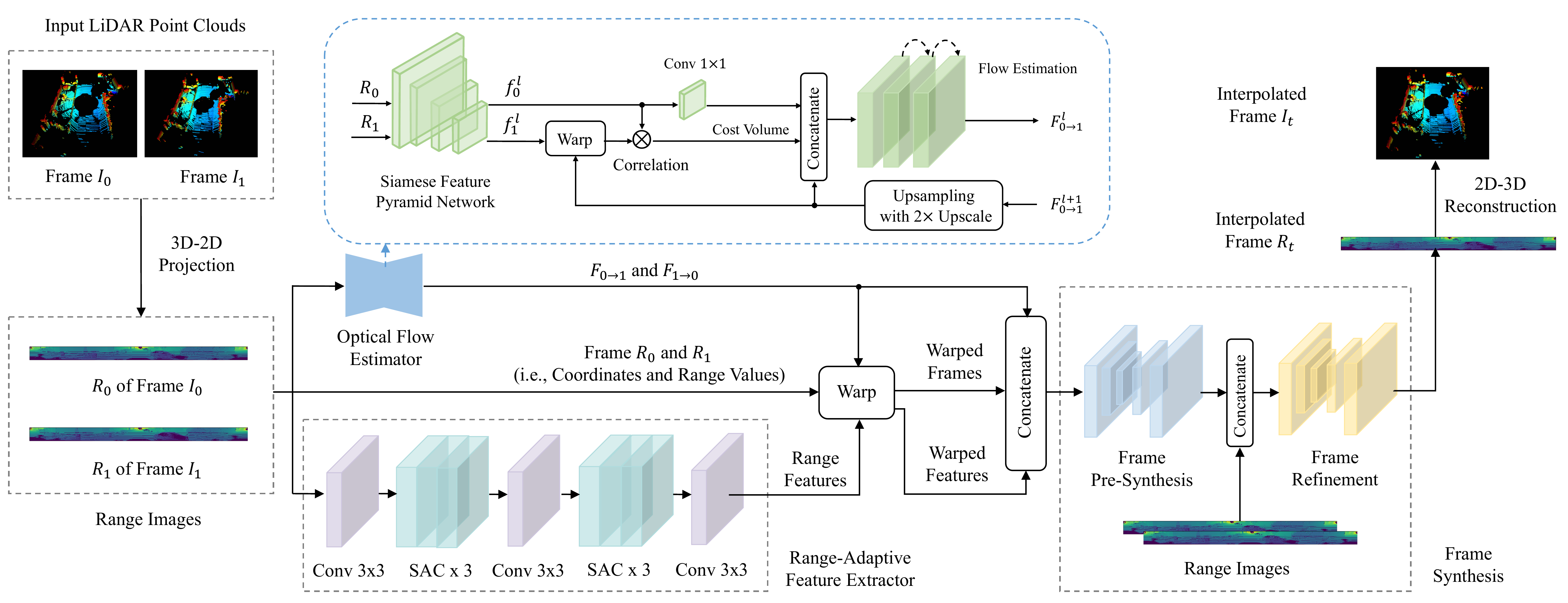}
	\caption{\textbf{The flowchart of the proposed range-adaptive LiDAR point cloud frame interpolation method.} It mainly consists of the following modules: 3D-to-2D projection, the optical flow estimator, range-adaptive feature extractor, the wraping layer via softmax, frame synthesis, and 2D-to-3D reconstruction. Note that the blue dotted box represents the specific structure of our optical flow estimator, which is modified from PWC-Lite~\cite{liu2020learning}, a lightweight architecture of PWC-Net~\cite{sun2018pwc}.}
	\label{fig:struc}
\end{figure*}

Given two input frames $\mathbf{I}_{0}$ and $\mathbf{I}_{1}$ of 3D point clouds, our approach aims to synthesize an intermediate frame $\mathbf{I}_{t}$ at time $t \in[0,1]$. Firstly, a 3D-to-2D projection is conducted to generate two 2D range images $\mathbf{R}_{0}$ and $\mathbf{R}_{1}$ as the input frames of our model. Then, the bi-directional optical flows, denoted by $\mathbf{F}_{0 \rightarrow 1}$ and $\mathbf{F}_{1 \rightarrow 0}$ respectively, are estimated. Next, based on the optical flows, we warp the input frames and range features, with a warping layer via softmax splatting. After that, we employ frame synthesis networks to produce the interpolated frame $\mathbf{R}_{t}$ in the range-image format. Finally, the interpolated point cloud frame $\mathbf{I}_{t}$ is reconstructed from $\mathbf{R}_{t}$. Fig.~\ref{fig:struc} provides an overview of our framework. In what follows, the various stages of our framework will be respectively described in detail.

\begin{figure}[t]
	\centering
	\includegraphics[width=80mm]{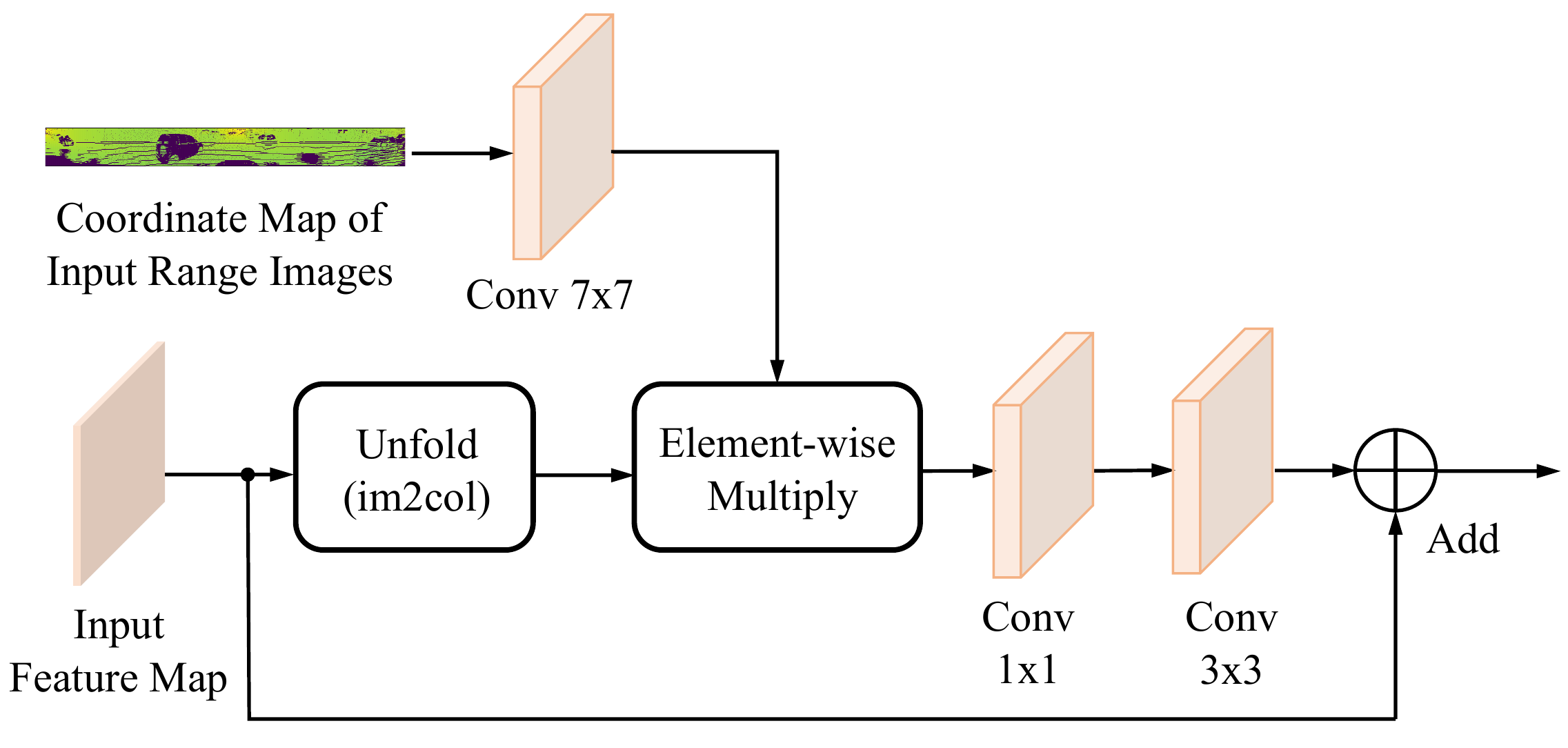}
	\caption{Structure of the spatially adaptive convolution (SAC), which is extracted from the work~\cite{xu2020squeezesegv3}. It is incorporated into our feature extractor to generate range features adaptively for range images.}
	\label{fig:rac}
\end{figure}

\subsection{3-D to 2-D Projection for Point Clouds}
In our pipeline, LiDAR point clouds are firstly projected into a set of 2D range images [one can see Fig.~\ref{fig:example} (c)], which is defined as~\cite{RangeNet}

\begin{equation}
	\left(\begin{array}{c}
u \\v
\end{array}\right)=\left(\begin{array}{c}
\frac{1}{2}\left[1-\arctan (y, x) \pi^{-1}\right] \quad w \\
{\left[1-\left(\arcsin \left(z r^{-1}\right)+\mathrm{\theta}_{\mathrm{up}}\right) \mathrm{\theta}^{-1}\right] h}
\end{array}\right)
\label{e1}
\end{equation}
where $(x, y, z)$ are the 3D coordinates of a certain point in point clouds, and $(u, v)$ are the 2D coordinates of the corresponding pixel after the projection. $(h, w)$ are the height and width of the desired projected range image. $\theta=\theta_{u p}+\theta_{\text {down}}$ is the field-of-view of the LiDAR sensor along its vertical direction, and $r=\sqrt{\left(x^{2}+y^{2}+z^{2}\right)}$ is the range from the point to the center of the sensor's coordinate system. In this way, each 3D point $(x, y, z)$ is projected into a pixel $(u, v)$ of the range image. 

\subsection{The Proposed Frame Interpolation Model}
Our model is composed of the following sub-modules: the optical flow estimator, the range-adaptive feature extractor, and frame synthesis networks. We will subsequently discuss its individual components.

\vspace{0.3em}
\noindent \textbf{Optical flow estimator.}
It is well-known that PWC-Net~\cite{sun2018pwc} is one of the state-of-the-art optical flow methods. However, as a sub-module in the model of point cloud frame interpolation, its computational complexity is not desirable. Therefore, we modify PWC-Lite~\cite{liu2020learning}, a lightweight architecture of PWC-Net, as our flow estimator. Our method follows the main pipeline of PWC-Lite, but with some adaptions in two aspects. (1) The original implementation aims to predict bi-directional optical flows for multiple frames. For low complexity and consistency with the input frame numbers of our system, we reduce the model size to estimate the optical flow only for two frames. The structure is depicted in Fig.~\ref{fig:struc} (top). (2) The loss function adopted in original PWC-Lite, is a measurement of pixel-wise similarities between color images. However, as mentioned previously, the inherited characteristics of range images are drastically different from that of color images. Therefore, Charbonnier penalty function~\cite{bao2019depth} is utilized as the loss function due to its smoothness and robustness. 

To further show the effectiveness of our modifications, we additionally fine-tune PWC-Net for our frame interpolation method. The specific results can be seen in Table~\ref{optical}. It can be obviously found that for range images, our flow estimator provides a better performance.

\vspace{0.3em}
\noindent \textbf{Range-adaptive feature extractor.} 
In~\cite{xu2020squeezesegv3}, a spatially-adaptive convolutions (SAC) is adopted to achieve more efficient point cloud segmentation. Intuitively, it also could be a powerful tool for our feature extractor for range images. Therefore, we extract the SAC module (Fig.~\ref{fig:rac}) to incorporate it into our feature extractor. As shown in Fig.~\ref{fig:struc} (bottom), we adopt the convolutional layer with the $3\times3$ kernel followed by three SAC modules, where Leaky ReLu ($\alpha$ $=$ 0.1) is utilized after each convolutional layer. Note that there is not any normalization layer, e.g., batch normalization, since it has been proven that the batch normalization is not suited for image quality improvement~\cite{8099518}. We jointly train the feature extraction network with our entire model, which learns range features adaptively for point cloud frame interpolation.

\vspace{0.3em}
\noindent \textbf{Warping layer.}
It has been shown that forward warping via softmax splatting delivers new state-of-the-art results for video frame interpolation~\cite{Softmax}. It can effectively address the issue incurred by the case that multiple source pixels are mapped to the same target location. Hence, in our work, with the estimated bi-directional optical flows, softmax splatting is employed to warp the input frames (i.e., containing $x$, $y$, and $z$ coordinates, and range $r$), and the features of range images.

\vspace{0.3em}
\noindent \textbf{Frame synthesis.}
To generate the intermediate range image, following the coarse-to-fine architecture, we design the frame synthesis networks consisting of two cascaded U-Nets~\cite{U-Net} with the same architecture, i.e., the frame pre-synthesis network and the frame refinement network respectively in Fig.~\ref{fig:struc}. This is motivated by the fact that, benefited from a skip connection between the encoder and decoder, U-Net is well suitable for the operation at pixel level~\cite{Softmax}. To reduce the computation complexity, the cascade U-Nets are modified with less depths, which contain a six-layered encoder and five-layered decoder. We use $3\times3$ kernels for all convolutional layers, followed with Leaky ReLu ($\alpha$ $=$ 0.1). It should be noted that as illustrated previously, batch normalization is not adopted. 

Note that the output of the frame pre-synthesis network is a `coarse' version of the range image. To yield one `fine' frame with better details, we concatenate it with two original range images as guidance information, and fed them into the frame refinement network to generate the interpolated range image $\mathbf{R}_{t}$. Finally, the interpolated point cloud frame $\mathbf{I}_{t}$ can be reconstructed from $\mathbf{R}_{t}$ based on Equation~(\ref{e1}). It is also worth noting that our model can generate the arbitrary intermediate frame $\mathbf{I}_{t}$ at time $t \in[0,1]$.

\subsection{Implementation Details}
\label{ID}
\textbf{Loss function.}
In this work, we train the flow estimator and the frame interpolation model, both by optimizing the Charbonnier penalty function~\cite{charbonnier1994two}, i.e., $f(x)=\sqrt{x^{2}+\varepsilon^{2}}$ ($\varepsilon = 1e-6$), but with different terms respectively. 

For the flow estimator, suppose in each layer $i$ of the feature pyramid network, $R_{i}$ is the ground-truth frame, and $\widehat{R}_{i}$ is the target frame, which is obtained by backwarping input frame with predicted optical flow $\mathbf{F}_{0 \rightarrow 1}$. Then the loss function $\mathcal{L}_{1}$ can be computed by

\begin{equation}
\mathcal{L}_{1}(\widehat{R}, R)=\frac{1}{L} \sum_{i=1}^{L}\sum_{x} f(\widehat{R}_{i}(x)-R_{i}(x)),
\end{equation}
where $L$ is the total number of pyramid levels. 

For our interpolation model, the loss function contains two loss terms, where the first term is for the pre-synthesis network, and the second term is for the frame refinement network. Assume $I$ is the ground-truth frame, and $\widehat{I}$ is the final synthesized frame, while $\widehat{I}_{p}$ is the corresponding frame from the frame pre-synthesis network. The loss function $\mathcal{L}_{2}$ can be formulated as

\begin{equation}
\mathcal{L}_{2}(\widehat{I}, I) = \lambda_{1} \sum_{x} f(\widehat{I}_{p}(x)-I(x)) + \lambda_{2}\sum_{x} f(\widehat{I}(x)-I(x)),
\end{equation}
 where we empirically set $\lambda_{1}=0.1$ and $\lambda_{2}=1$. The rationale behind this setting is that `fine' frame interpolation, as the final objective, is always dominant, while the effect of `coarse' version generation is limited.
 
\vspace{0.3em}
\noindent \textbf{Training dataset.}
We conduct our training process on the KITTI dataset~\cite{Geiger2012CVPR}. It provides the odometry benchmark containing 22 sequences of point cloud data, which are collected by LiDAR sensors Velodyne HDL-64E, covering various road scenes. For each sequence, we partition every 3 consecutive frames into a triplet. With regard to each triplet, the first and third frames are as the input frames of our model, while the second frame is the ground-truth frame. In this way, 16 sequences (10327 triplets) are chosen as the training dataset. Among the others, 3 sequences (2400 triplets) are for validation dataset, while 3 sequences (1784 triplets) are for test dataset. It is worthwhile to noting that each dataset covers all kinds of scenes.

\vspace{0.3em}
\noindent \textbf{Training strategy.}
Note that PWC-Lite is designed for videos (i.e., a sequence of color images), while the range images have diverse characteristics from color images. Therefore, instead of using the parameters of a pre-trained PWC-Lite, we train our flow estimator from scratch for 40 epochs to converge. The ADAM optimizer is adopted with $\beta_{1}=0.9$, $\beta_{2}=0.999$, and $\varepsilon=1e-7$. The initial learning rate is set to $1e-4$ and reduced by 0.1 when the validation loss has stopped decreasing for 4 epochs. 

After obtaining the well-trained flow estimation model, its parameters will be freezed. Then, we train the interpolation model for 40 epochs, with the same initial learning rate and learning rate schedule as mentioned above. Next, we reduce the learning rate to $1e-5$ and fine-tune the entire model for another 20 epochs. Similarly, the learning rate during fine-tuning process will be decayed by a factor of 0.2, when the validation loss has stopped declining for 5 epochs. The ADAM optimizer is with the default setting and batch size is set to 1. We implement all the training process with the NVIDIA 1080Ti GPU, which takes about 4 days to converge.

\begin{figure*}[t]
	\centering
	\includegraphics[width=175mm]{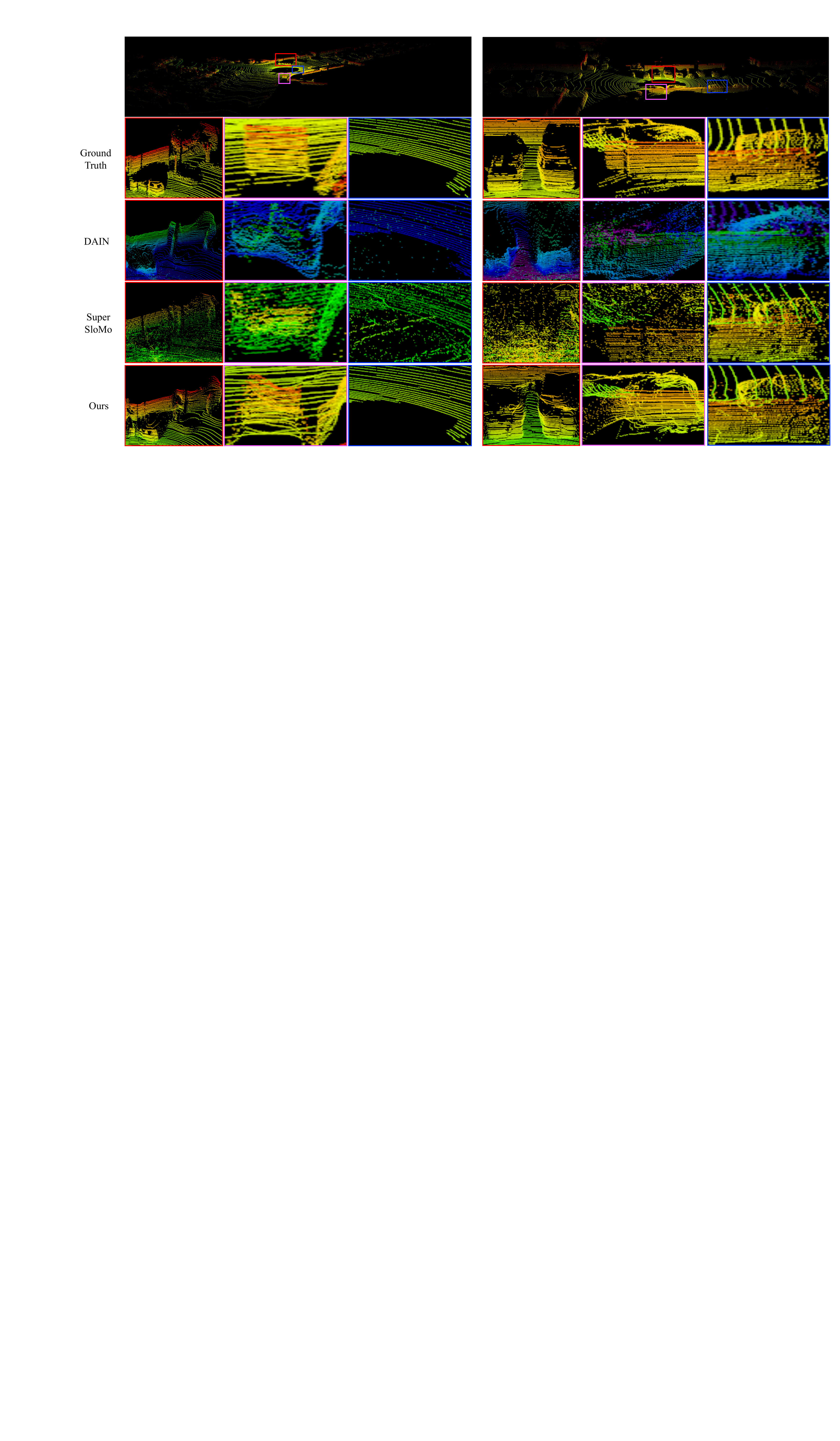}
	\caption{\textbf{Visual comparisons for two frames on KITTI dataset with zoom-out and zoom-in views.} For all the shown examples, especially the frames with edges or points lie close to straight lines, our method perceptually outperforms other state-of-the-arts by a large margin.}
	\label{fig:ex}
\end{figure*}

\section{Experimental Results}
We evaluate our method on the test dataset derived from the KITTI dataset, and compare it to state-of-the-art video frame interpolation methods quantitatively. The same approaches for 3D-2D projection and 2D-3D reconstruction using Equation~(\ref{e1}), are applied into the video frame interpolation methods under comparison, where we set $H=64$ and $W=2048$, as used in~\cite{RangeNet,xu2020squeezesegv3}. All experiments are implemented using PyTorch and run on the same machine using an Intel Core i7-7700K and 11G RAM with GTX 1080Ti. 

\subsection{Evaluation Dataset and Metric}
\vspace{0.3em}

\noindent \textbf{Test dataset.}
As described in Section~\ref{ID}, three sequences of the KITTI dataset are chosen as the test dataset, containing totally 1784 triplets. We randomly select a quarter of the test dataset for this evaluation. 

\vspace{0.3em}
\noindent \textbf{Metric.}
It should be noted that metrics adopted in video frame interpolation, i.e., PSNR and SSIM, are for 2D normal images, which are obviously not proper for 3D LiDAR point clouds. It is well-known that \emph{Symmetric Nearest Neighbor Root Mean Square Error} (SNNRMSE)~\cite{6907779}, is typically used to evaluate the quality of point cloud reconstruction. Therefore, in this paper, it is utilized for the performance evaluations of point cloud frame interpolation. 

Assume $P$ and $Q$ are the original frames and interpolated frames of point clouds, where each point $p$ in $P$ can be find its closest point $q$ in $Q$ by the k-d tree search, denoted as $q=f(p, Q)$. Then, the \emph{mean squared error} (MSE) is obtained by measuring the euclidean distance of each point p in P to its nearest neighbor q in Q. That is,

\begin{equation}
	\operatorname{MSE}(P, Q)=\sum_{p \in P}(p-q)^{2} /|P|,
\end{equation}
where $|P|$ is the point number of $P$. Similarly, $\operatorname{MSE}(Q, P)$ can be computed as well. Lastly, SNNRMSE is obtained by

\begin{equation}
	\operatorname{RMSE}_{SNN}(P, Q)=\sqrt{\frac{MSE(I, Q)+MSE(Q, I)}{2}}.
\end{equation}

\subsection{Performance Evaluations}

\vspace{0.3em}
\noindent \textbf{Evaluation for flow estimation.}
As illustrated previously, our flow estimator is modified from PWC-Lite~\cite{liu2020learning}, which is a lightweight model of the state-of-the-art method PWC-Net~\cite{sun2018pwc}. To show the effectiveness of our adaptations, we compare our flow estimator with these two approaches. Since our work aims at frame interpolation issue, we just make comparisons by the inference loss via Equation (2) for reference, during the designing stage of our flow-estimator. It is worth to mention that we also provide the result of the fine-tuned PWC-Net, which outperforms the original PWC-Net. It turns out that the original models for flow estimation are trained for color images. Therefore, for the final evaluations of our entire model, instead of using the weights of pre-trained models under comparison, we fine-tune all the methods using our training dataset.

As shown in Table~\ref{optical}, our proposed method delivers the best results on flow estimation in all key aspects, i.e., runtime and inference loss. Another illustration for the effectiveness of our flow estimator can be found in Table~\ref{ex}, where the fine-tuned PWC-Net is adopted as the flow estimator in our model for comparison, denoted as `Ours + PWC-Net-ft '.

\begin{table}[t]
\centering
\small
\caption{\textbf{Quantitative comparisons for flow estimation methods on the KITTI dataset.} `-ft' means fine-tuning on the training set. The proposed flow estimation method outperforms other approaches in terms of loss and runtime.}
\vspace{6pt}
\begin{tabular}{ccccc}
\hline
&Method          & \multicolumn{1}{c}{Runtime (s)} & \multicolumn{1}{c}{Loss} &            \\
\hline
&PWC-Net ~\cite{sun2018pwc}   & 0.0177                      & 3.1319              &                 \\
&PWC-Net-ft & 0.0183                      & 0.7226                &               \\
&PWC-Lite~\cite{liu2020learning}        & 0.0143                      & 3.1709                   &            \\
\hline
&Ours            & \textbf{0.0142}                      &  \textbf{0.5242}  & \\
\hline
\end{tabular}
\label{optical}
\end{table}

\begin{table}[t]
\centering
\small
\caption{\textbf{Quantitative comparisons for frame interpolation methods on the KITTI dataset.} `-ft' means fine-tuning on the training set. The proposed method outperforms other approaches in RMSE with real-time performance.}
\vspace{4pt}
\begin{tabular}{cccc}
\hline 
Method     & Runtime (s) & RMSE (m)     &  \\
\hline
Super SloMo-ft~\cite{jiang2018super} & \textbf{0.0269} & 0.5393   &    \\
DAIN-ft ~\cite{bao2019depth}  & 0.2380  & 0.6224   &    \\
Ours + PWC-Net-ft ~\cite{sun2018pwc}   & 0.0976 & 0.3684   &    \\
\hline
Ours      & 0.0961 & \textbf{0.3107} &  \\
\hline
\end{tabular}
\label{ex}
\end{table}

\vspace{0.3em}
\noindent \textbf{Evaluation for frame interpolation.}
To justify the effectiveness of the proposed frame interpolation method, we compare our method with the other two state-of-the-art methods for video frame interpolation, i.e., DAIN~\cite{bao2019depth} and Super SloMo~\cite{jiang2018super}. Table~\ref{ex} reports the results on the test dataset. It can be clearly observed that our method outperforms other approaches in term of RMSE. Though our approach is computationally much more expensive than Super SloMo, we can still deliver the real-time performance (running faster than the frequency of LiDAR sensors, i.e., 10Hz).

In addition, Fig.~\ref{fig:ex} shows two representative frames of interpolated samples. It can be seen that our method is able to constantly provide the most desirable interpolation results from the viewpoint of perceptual quality, especially on the regions with edges or points lie close to straight lines (referred to red and blue-framed regions cropped from the frames respectively).

\section{Conclusion}
In this paper, we propose a novel framework for LiDAR point cloud frame interpolation, which can be applied to reduce the amounts of point cloud transmission. The main contribution lies in the newly developed model, which exploits spatial-temporal features, extracted from 2D representations (range images) of 3D point clouds, as the auxiliary information to synthesize the intermediate frame. Considering that inherited characteristics of range images are diverse from that of normal color images, we introduce spatially adaptive convolutions to generate range features adaptively, while a high-efficient flow estimator is presented for optical flow extraction. Lastly, we warp input frames and range features with estimated flow to synthesize the intermediate frame. Extensive experiments have demonstrated that the proposed method can consistently deliver superior results and the most visually-pleasant quality, compared with that by using other state-of-the-art video frame interpolation methods. The directions for out future work are to develop a faster and more efficient model, and integrate the model into multiple point cloud compression structures.

\bibliographystyle{IEEEbib}
\bibliography{reference}

\end{document}